\documentclass[12pt]{article}
\usepackage{titlesec}
\usepackage{titling}
\usepackage{amsmath, amsthm, epsfig, amssymb}
\usepackage{hyperref}
\usepackage[font={small,sf},labelfont={small,sf,bf}]{caption}
\usepackage[normalem]{ulem}

\usepackage{authblk}
\usepackage{verbatim}
\usepackage[titles]{tocloft}

\usepackage{epigraph}
\usepackage[utf8]{inputenc}
\usepackage{cleveref}
\crefname{section}{§}{§§}

\renewcommand{\th}{\theta}

\usepackage{float}

\usepackage{enumitem}
\setlist[itemize]{leftmargin=*}
\setlist[itemize]{itemsep=-1ex}
\setlist[itemize]{topsep=0ex}

\title{{\normalfont\Large Explanatory models in neuroscience: \vspace{-5pt}
\linebreak Part 2 -- Constraint-based intelligibility
}\vspace{-15pt}}

\author{Rosa Cao and Daniel Yamins\thanks{Department of Philosophy (RC); Departments of Psychology \& Computer Science (DY); and Wu Tsai Neurosciences Institute (RC \& DY), Stanford University.}\vspace{-75pt}}

\date{}

\begin{document}

\maketitle

\setlength{\cftbeforesecskip}{3pt}
\setcounter{tocdepth}{1}
\tableofcontents

\thispagestyle{empty}


\begin{abstract}
Computational modeling plays an increasingly important role in neuroscience, highlighting the philosophical question of how computational models explain. In the context of neural network models for neuroscience, concerns have been raised about model intelligibility, and how they relate (if at all) to what is found in the brain.  We claim that what makes a system intelligible is an understanding of the dependencies between its behavior and the factors that are causally responsible for that behavior. In biological systems, many of these dependencies are naturally ``top-down'': ethological imperatives interact with evolutionary and developmental constraints under natural selection. We describe how the optimization techniques used to construct neural network models capture some key aspects of these dependencies, and thus help explain \emph{why} brain systems are as they are -- because when a challenging ecologically-relevant goal is shared by a neural network and the brain, it places tight constraints on the possible mechanisms exhibited in both kinds of systems. The presence and strength of these constraints explain why some outcomes are more likely than others. By combining two familiar modes of explanation -- one based on bottom-up mechanism (whose relation to neural network models we address in a companion paper) and the other on top-down constraints, these models can illuminate brain function.
\end{abstract}


\epigraph{\textit{Nothing in biology makes sense except in light of evolution.}}{--- Theodosius Dobzhansky}

\epigraph{\textit{Nothing in neurobiology makes sense except in light of behavior.}}{--- Gordon Shepherd}

\section{Introduction}
\label{sec:intro}

In the companion paper on mechanisms,\footnote{See \url{https://arxiv.org/abs/2104.01490}.} we suggested that an explanation should illuminate the \textit{dependencies} between the phenomenon of interest and the factors involved in generating it.  An intuitive way for a model to be ``explanatory'' for a target natural system requires that it:
\begin{enumerate}
    \item be a \textit{mechanistic} account of causal relationships in that target system, and 
    \item be \textit{intelligible} in the sense of being \textit{cognitively manipulable}
\end{enumerate}

Having discussed mechanism in the companion paper, we turn now to intelligibility. We think that what makes a system intelligible is an understanding of the dependencies between its behavior and the factors that are responsible for that behavior. These factors may be bottom-up - such as the mechanistic components and organization already discussed.  But in biological systems, they can also be top-down: there can be evolutionary goals and historical or developmental constraints that are responsible for shaping a system. An explanation of \textit{why} a system is the way that it is can then appeal to this broader range of dependencies.

Cognitive manipulability is a way of capturing the aspect of explanation that depends on \textit{us}, on our capacity for understanding the explanations that we find, and more specifically, the models that we build.  We want our explanations to be the kinds of things whose import is cognitively \textit{available} to us -- whether directly (e.g. through visualization) or indirectly, through a suite of tools that allow us to efficiently query the explanation.  So for example, we might make salient the dependencies that our model embodies by manipulating some aspects of the model to see the effects on others. These aspects could be mechanistic, or they could be top-down constraints which shaped the model, making it turn out the way it did.

In straightforward explanatory situations, the relationship between the mechanistic facts about a system and what that system does are immediately apparent, especially if the system is simple, in the sense of having either relatively few parts, or relatively few \textit{kinds} of parts or kinds of interactions. In these cases, a classic path to intelligibility is via concise mathematical description of a system's dynamics that allows us to "eyeball" its behavior under various situations without having to actually perform all the calculations necessary for precise prediction.

When the system is mechanistically complex, by contrast, a way of making an explanation tractable is to find some intermediate level of description which would allow us to capture the relevant dependencies, organizing the lower-level implementation facts in order to highlight their relevance to producing the phenomenon we are interested in explaining (in systems neuroscience, perhaps behavior or performance on a task). This intermediate level of description is often given in terms of \textit{function}, where the idea is that the phenomenon of interest admits of a functional decomposition into simpler sub-tasks, and we can (hope to) make sense of decompositions of the system into subsystems that perform those sub-tasks.

One familiar way that this is done is by appeal to \textit{representations} in neuroscience. For example, to say that cells in cortical area V1 (more on this brain area below) represent \textit{edges} makes V1 seem intelligible, because we can see \textit{how} being responsive to edges can help the visual system recognize objects. We link particular neural activity to particular stimuli, by saying that the \textit{function} of that activity is to represent that stimulus. As Marr noted, to speak in terms of (computational) function is to say something about \textit{why} certain features of a mechanism are present.

Sometimes even this clear functional decomposition may not be available however. In such cases, the explanatory project is more difficult, but can still gain some purchase by looking at what the system as a whole was selected or optimized to do. This allows us to illuminate another kind of dependency, between the constraints that had to be met, and the solution that arose as a result of the interplay between those constraints and the resources available to the system.

Instead of gaining traction by simplifying the ``how'' of a mechanism (as simple mathematical formulae do), evolutionary explanations provide answers to ``why'' questions. So for example: Why does the system have \emph{this} form? Answer: Because it had to perform \emph{that} function~\cite{dennett1996darwin} -- and, importantly, the world is such that having to perform a particular function rules out many other forms. A model which articulates the design constraints in an evolved system thus provides another route to intelligibility. 

By evaluating how \textit{constraining} a function or task is, we can reason about \textit{necessary} (or at least \textit{likely) }conditions on any mechanism that successfully performs it. This speaks to the intuition that to explain a phenomenon is not just to \textit{describe} it, but also to say why it should have happened \textit{this} way, rather than some \textit{other} way -- or why what we observe is more \textit{likely} than the alternatives, perhaps even \textit{necessary}. This can happen when the constraints are strong -- and we will want to say more about what makes for strong constraints.

In the actual case of neural networks (NNs), it may turn out that no efficient encapsulations of the relevant dependencies are available -- either of how the system's behavior depends on inputs, or how its behavior changes in response to perturbations of the mechanism. We think that it is primarily this apparent characteristic of NN models that provokes skeptics to say that they are unintelligible. 

We agree with critics that NN models (whether biological or artificial) do not (yet?) yield to succinct mathematical formulae for their future states that scientists can estimate in their heads. However, we will show how, in least in some circumstances, they do admit of this second form of intelligibility that is distinctively biological, derived from reasoning about the constraints faced by a functional system - a system that is the way it is for a \textit{purpose} or a \textit{reason}. That is, we can probe and understand dependencies between the solutions to ethological problems, and the constraints involved in generating them.

We will start by examining the Olshausen and Field model of early visual cortex, as an example of how much can be derived from constraints on a system.  We'll then move to the more complex modern hierarchical convolutional NN models. Again, constraint-based reasoning can help illuminate \emph{why} they have the particular features that they do. These are particular examples of interest, but they follow from some more general principles: that though models may appear quite different on the surface, it is their shared features that will \textit{explain} why they are successful at prediction, \textit{when the predictions are difficult}. The shared features we focus on are, again, the constraints involved in building both the target system and the model of it. We conclude by drawing some connections to the constraints manifest in evolutionary landscapes, the kinds of explanation that they license, and the kinds of intelligibility that they provide.

\begin{figure}
\centering
\includegraphics [width=.95\linewidth]{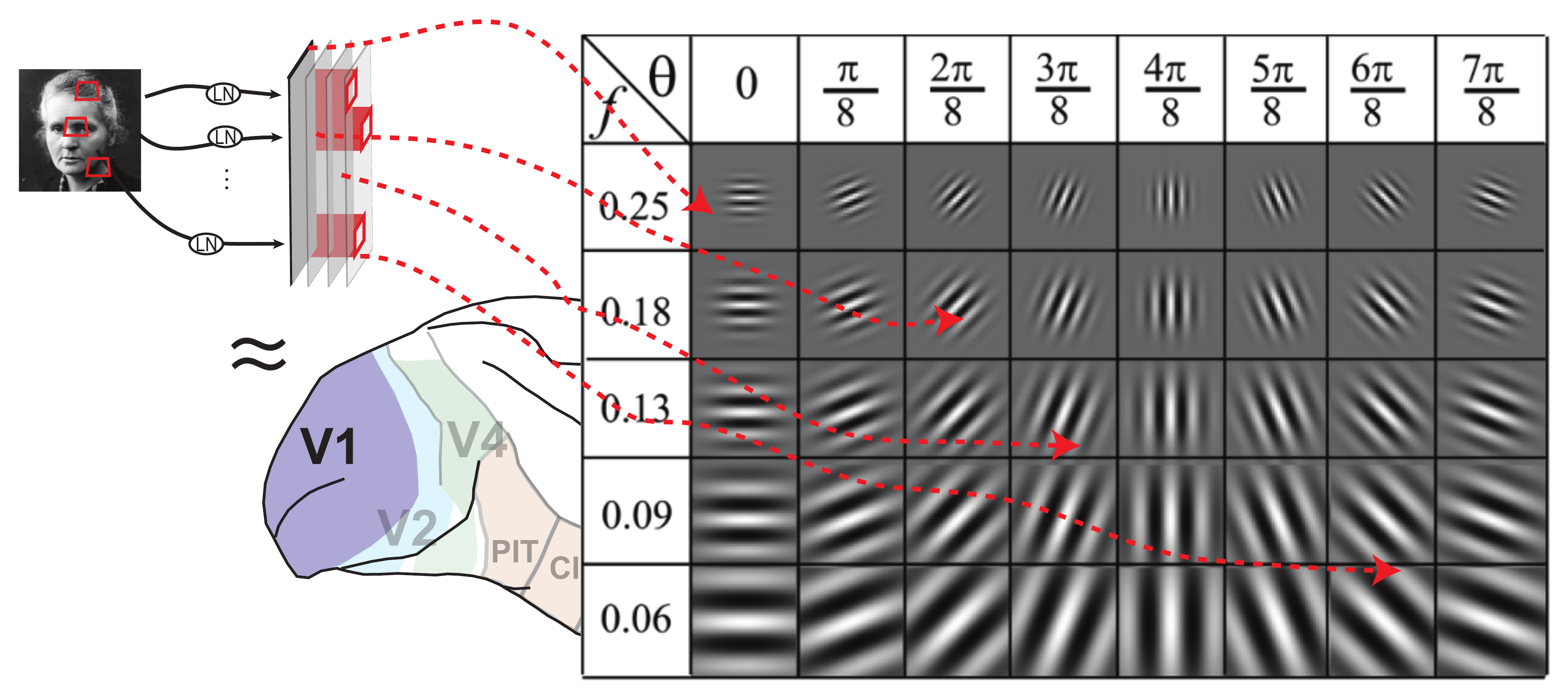}
\vspace{-3mm}
\caption{\textbf{Gabor-filterbank convolution model of V1 cortex.} Both V1 and an early layer of an HCNN can be understood as performing a similar set of computations: convolving their inputs with Gabor filters. The equation for any individual filter (the individual grey slices) is given by the Gabor wavelet equation, e.g. $g(x, y; f, \th) \sim exp \left [- \frac{x^2 + y^2}{2} \right] exp \left[ 2 \pi i f(x \cos \th + y \sin \th)\right]$, where $x,y$ are the coordinates in the input image and $f$ and $\th$ are, respectively, the frequency and orientation of the edge. The full filterbank contains frequencies $f$ and orientations $\th$ covering a range useful for extracting edges from natural images. 
\label{fig:v1}}
\vspace{-3mm}
\end{figure}

\section{Beyond Classic Bottom-Up Intelligibility}
Being able to see how choices of the components in a model affect functional outcomes is a key ingredient of functional intelligibility.  Achieving such understanding is perhaps easiest when there is a simple structural equation for the model, considered as an arrangement of its components. 
As mentioned in \S2 of the companion paper to this one,\footnote{See \url{https://arxiv.org/abs/2104.01490}.} perhaps the most celebrated set of findings in visual neuroscience characterizes neurons in early cortical area V1 in essentially these terms. A line of experiments begun by Hubel and Wiesel~\cite{hubel1959receptive, hubel1962receptive} and carried further by many others~\cite{Carandini:2005kw, movshon1978spatial, ringach2002orientation} showed that V1 neurons can at least in part be described as responding optimally to environmental ``edges'' of varying frequencies and orientations (see Fig. \ref{fig:v1}).
Over the course of several decades, it became clear that Hubel and Wiesel's array of `edge-detecting' V1 neurons could be re-described in a mathematically compact way: as performing spatial convolution with the so-called ``Gabor wavelet filterbank''.
Models convolving input images with the Gabor filterbank, followed by simple rectification and normalization nonlinearities, have achieved striking success in characterizing V1 neural responses~\cite{Carandini:2005kw, ringach2002orientation, willmore2008berkeley}. 
While this characterization turns out to be a bit too simple given what we now know~\cite{cadena2019deep}, it is nonetheless a good enough approximation to serve many useful predictive purposes. 

The key point is that Gabor filters are generated by a very simple closed-form equation:
\begin{equation}
\label{eq:gabor}
g(x, y; f, \th) \sim exp \left [- \frac{x^2 + y^2}{2} \right] exp \left[ 2 \pi i f(x \cos \th + y \sin \th)\right],
\end{equation}
where $x,y$ are the coordinates in the input image and $f$ and $\th$ are, respectively, the frequency and orientation of the edge. 
The full filterbank contains frequencies $f$ and orientations $\th$ covering a range useful for extracting edges from natural images.
This formula does specify a runnable model in the sense of the previous section. But what makes it seem intelligible is how directly cognitively accessible it makes the consequences of various manipulations, how \textit{apparent} it makes the dependencies between mechanism and behavior.
For example, it is clear from the mathematical structure of eq. \ref{eq:gabor}, \emph{without} having to actually run the model, how changing the filterbank by adding or subtracting given orientations will improve or degrade edge detection performance. Moreover, it is possible to \textit{visualize} the qualitative activation response to particular stimuli, e.g. how oriented bars of a given orientation excite units of nearby preferred orientations.

In contrast, with deep neural networks, the very depth of the network makes it hard to visualize (unaided) what the outputs of the network will be for any given stimulus.  Moreover, there is no guarantee that we will ever be able to find a short formula for the final parameter choice of the network: the optima might be very hard to describe in terms of the kind of short formula that characterizes a Gabor filterbank. Indeed, there is no good reason to expect the existence of such a short formula for the parameters in a real biological system itself: any system that is the result of an iterative optimization process is likely to have a complex form.

Fortunately, a simple formula is not the only way in which it is possible to achieve some level of cognitive tractability.
In biological systems, we often appeal to the idea of \textit{function} in the sense of identifying what some trait or system is \textit{for}.  Functional explanations answer \textit{why}-questions, connecting the mechanism of system to what it must do in order to promote the survival or fitness of the organism as a whole.  These are in a sense top-down constraints on the system, and so we call this a top-down approach to functional intelligibility. We try to determine whether a fitness-enhancing behavior places strong restrictions on the types of possible solutions (e.g. neural configurations) that will produce it, and what (if any) the common characteristics of all such solutions are. 

In the pioneering work of Olshausen and Field~\cite{olshausen1996emergence}, it was shown that V1-like Gabor wavelet patterns could arise in a single-layer convolutional neural network that is optimized to transmit its input while minimizing the amount of overall activation needed to do so. QUalitatively similar patterns to Gabor wavelets arise from this compressive ``sparse-autoencoding'' optimization, without having to be built in.

The sparse autoencoder model of early visual cortex does not yield a simple formula (and in fact what arises from the optimization is not precisely the Gabor wavelet formulation, despite bearing a passing resemblance), but it does yield intelligibility in another way: the structure of the objective function (in particular, its two competing goals of representation accuracy and activation efficiency) provide a useful framework in which to test explanations that appeal to the evolutionary pressures shaping V1.

We can think of the HCNN models as a further step along this route, but with a harder task goal, and a more fully articulated architecture -- and correspondingly, the ability to provide ``why'' explanations not only for early cortical areas like V1 but also intermediate and higher areas like V4 and IT. In fact, it turns out that V1 can also be modeled as an early layer of the \emph{same} deep HCNN model of the visual system whose intermediate and higher layers can used to describe intermediate and higher visual cortical areas~\cite{cadena2019deep}. Moreover, this HCNN description of V1 is substantially more predictively accurate with respect to the primate brain than either the short equation of the Gabor-wavelet model or the relatively simple optimization goal of the sparse-autoencoder model.  In other words, the neurons of V1 are \emph{neither} exactly a Gabor wavelet filterbank \textit{nor} the exact solution to a demand for efficient sparse representation. Instead, they are better understood as being constrained by the requirement that they serve as the ``front-end'' to a deeper system that, downstream, must solve high-level visual behavioral tasks such as object categorization.  This fuller description is still cognitively tractable along one dimension, since it provides answers to a very non-obvious ``why'' question about V1 neural responses, but is unfortunately quite a bit less tractable than the simple-equation-description given by the Gabor wavelet model. We believe this increase in cognitive cost accounts for widespread intuitions about the ``opacity'' or ``inexplicability'' of deep neural network models. However, it is not fatal to the project of understanding why the system works in a broader sense.

We will focus in the remaining sections of this paper on the top-down intelligibility that can be provided by a well-motivated optimization principle. We argue that goal-driven models of the ventral pathway are a prime example of how downstream behavior can place strong constraints on architecture.

\section{Goal-Optimized HCNNs Aren't Mere Curve-Fits} 
\label{subsec:top_down}

Optimization goals induce a high-dimensional landscape whose points represent different settings for the parameters of the system (Fig. \ref{fig:contravariance}).  Each position on that landscape can be evaluated for how well a system with those parameters performs at the task.  One way to make sense of why a system is the way it is is to see where it sits on this landscape, and how it got there, given where it started.\footnote{The landscape metaphor, originally introduced by Sewall Wright \cite{Wright:1932landscape}, has been widely used in the context of evolutionary explanations. For all its flaws and limitations in the context of theorizing about biological evolution, it remains a useful heuristic for our purposes here.\cite{Kaplan:2008landscape, Calcott:2008landscape, GAVRILETS1997307}}

Thinking about the case of primate vision, we know that it is possible to obtain \emph{at least one} solution to the goal of extracting behaviorally relevant information from visual stimuli.  But we also want to know: how many such solutions exist? How similar are any two solutions, and what ways are they similar (and different)?  

At one extreme, we might imagine discovering that there is only one peak in the landscape.  That is, there is a unique best solution to the task, given the architecture class, and moreover, pretty good solutions are all similar to the best solution and to each other.  This is to learn something important about any system that performs the task: namely, that it \textit{had} to have the features that it did, in order to do what it does.  Inevitability, when it obtains, is a good feature of explanation -- it is enlightening to discover that an outcome was not an accident, but the necessary outcome of some confluence of factors.

At the other extreme, it may be that there are many peaks (solutions), and that they are not particularly close to each other.  That would make us think that the outcome we observe, whether in a natural or an artificial system, is mostly an accident of history.  Things could have been otherwise, and no structural facts (about the task or the parameters) forced any particular observed outcome.  We can conclude only that the features of the system are sufficient to produce the outcome. 

Depending on how many parameters and hyperparameters we allow to vary, the landscapes themselves will be expressive of more or fewer possibilities.  But once those are fixed (that is, once we commit to a particular architecture-type), the goal will determine the shape of the landscape.  With this in mind, we can see that we should strive for ethologically realistic tasks, to be performed by plausible architectures, when constructing models that aim to mirror the relationship between form and function exemplified by a biological target. For a goal-optimization approach to succeed in generating biologically relevant models, it must be the case that performance on top-down goals places a substantive constraint on network parameters for reproducing neural response patterns.

That is why the parameters of HCNN models for vision (e.g. in \cite{yamins:pnas2014} were optimized to solve a visual performance goal that is ethologically plausible for the organism, rather than being directly fit to neural data. This is the idea of \emph{goal-driven} modeling~\cite{yamins2016using}. We describe the relevant characteristics of these models in detail in the companion paper \cite{cao2021explanatory}, but the key finding is that these models provided good predictions of neural data despite never having been exposed to neural data -- a finding that should be quite surprising at face value.\footnote{In summary: modern HCNN models of the private visual cortex (such as the ones described in Yamins 2014) are well-defined computational programs accepting as input any image-like stimulus, and performing the ecologically-relevant task that is a proxy for the capacity of interest.  A deep HCNN is built with about the the right number of layers as there are observed brain areas in the ventral visual pathway. Then the parameters of the HCNN are optimized such that the resulting network is able to solve a challenging behavioral task -- typically, 1000-way object categorization in real-world images~\cite{Deng_3067}. Intriguingly, it turns out that the \emph{error patterns} of the outputs of these goal-optimized networks correspond to a large degree to those measured from data in human and primate behavioral experiments~\cite{rajalingham2018large}, even though the networks were optimized for overall performance rather than for producing any particular error pattern.  Most importantly for our purposes, it also turns out that HCNNs built this way are by a very long margin the best quantitative models of neural responses in every measured cortical area of the primate ventral visual pathway.

Model responses from hidden layers near the top of HCNNs are highly predictive of neural responses in IT cortex, up to linear transform, both in electrophysiological~\cite{yamins:pnas2014, cadieu2014deep}, and fMRI data~\cite{kriegeskorte:ploscb2014, gucclu2015deep}.  Importantly for our purposes, the HCNN models are also substantially better at predicting neural response variance in IT than ideal-observer semantic models which have perfect access to object category or other attributes~\cite{yamins:pnas2014}. Though the ideal-observer models ``solve'' the posited behavioral task (e.g. categorization) perfectly, they are \emph{not} in fact runnable models at all, and so are not constrained to generate the answer from only the real inputs that the system has (e.g pixels). The fact that the HCNNs, which \emph{are} runnable models, end up being much better predictors of neural responses shows that runnability is an important constraint on the system. 

Intermediate layers of the same HCNNs whose later layers match IT neurons also yield state-of-the-art predictions of neural responses in V4 cortex~\cite{yamins:pnas2014,gucclu2015deep}, the dominant cortical input to IT.  Again, the mapping between models and brain data uses a linear transform to perform the match. Similarly, recent models with especially good task performance have distinct layers clearly segregating late-intermediate visual area PIT neurons from downstream central IT (CIT) and AIT neurons~\cite{nayebi2018task}.  In other words, HCNN models suggest that the computations performed by the circuits V4 are structured \textit{in order that} that downstream computations in PIT and, subsequently, AIT, can support robust categorization in tasks that require the ability to deal with high-variation images.} 

These results are important because they suggest that high-level ecologically-relevant constraints on network function --- i.e. the categorization task imposed at the network's output layer --- really \textit{are} strong enough to shape upstream neural responses in a non-trivial way.

For the models of the visual system we have been discussing, the question was: are performance constraints imposed at the top layer of a network sufficiently strong to force units in hidden layers in the network to behave like real neurons in V1, V4 or IT? (Remember, no neural response data is used directly in training the HCNNs). This is an empirical question, and as it turns out, the neural networks resulting from this approach effectively model the biology as well or better than direct curve fits~\cite{cadena2019deep, yamins:pnas2014} to neural data.  In this case, combining\textbf{ two general biological constraints -- the behavioral constraint of object recognition performance, and the architectural constraint imposed by the HCNN model class} -- led to improved models of multiple areas through the visual pathway hierarchy.

Such a result suggests that we are closer to the first extreme described above, where solutions are similar to each other, or perhaps even unique, so that systems that succeed at the task are forced to have certain features. In this case, the features are exhibited by \textit{both} the primate visual system, and the artificial HCNN model of it. This is what explains the surprising result that HCNNs do a good job of predicting neural responses despite being optimized only for the visual task, and with no exposure to neural data.

It is a vaunted consequence of Maxwell’s equations that the speed of light “just falls out”. Those equations were derived from empirical fits to measured quantities, e.g. Faraday’s law, and it was not until much later that the relationship to the speed of light could be confirmed experimentally.  Nonetheless, that unlooked-for consequence serves as a kind of dramatic confirmation. When such surprising predictive successes occur, that gives us reason to believe that there are tight, perhaps \textit{necessary} connections between the things we fit directly and the things we accidentally ended up successfully predicting --- and this is just what we have in the neural network modeling results for the primate visual stream.

Maxwell's equations were \textit{not} just curve-fitting measurements about voltages and current: rather, they constituted a kind of explanatory model-building that captured broader dependencies among variables of interest.\footnote{For contrast, the neural network model of a L5PC described in \cite{Beniaguev:2019single} (and discussed in Section 4 of \cite{cao2021explanatory}) really is merely curve-fitting the biophysical properties of the cell.} So too with the goal-driven HCNN models described above. They seem to be operating in a regime where the dependencies between form and function are sufficiently strong, that to successfully fit one (the task goal) is also to learn about the other (the mechanistic structure of the target system). Thus we can answer a why-question: why does the ventral stream exhibit the response properties it does?  Because they are \textit{needed} for it to perform the task that it does.

\section{Difficult predictions and the no-miracles argument}

Critics have argued that neural network models are not really explaining the brain, but "merely curve-fitting the data" -- because in order to map model predictions to brain data, a fitting procedure needs to be run. We have discussed in the section on similarity mappings in the companion paper on mechanisms \cite{cao2021explanatory} why that procedure is no more curve-fitting (in this pejorative non-explanatory sense) than comparing the data from two different animals would be. However, another aspect of the complaint remains: that the models somehow manage to accurately-enough \textit{predict} brain data without yet providing any sort of \textit{explanation} of them.

In response, we can appeal again to the explanatory principle just described in the previous section, but running in the other direction: If the constraints (of getting behavior appropriate to the task, and the general architectural class) were not actually relevant, then it would be a kind of \textit{miracle} -- or wild coincidence -- for a model satisfying those constraints to somehow get the predictions for intermediate-level neural activity right.

The situation here similar to the one posed by the instrumentalist (or empiricist) challenge to scientific realism.  Sure, our scientific theories are on the whole successful. But history, the pessimistic meta-induction tells us, shows that theories successful in some limited domain of their initial application can still be false. And moreover, most past theories have been shown to be false, by our current lights.\cite{Laudan:1981pessimistic} So why think that success is any guide to truth?  Why think that our science is capturing anything real about how things are, rather than producing some temporary accommodation to the facts available at present?

However, the realist might ask in turn -- if our science isn’t capturing anything real about how things are, then \textit{why} is it successful?  It seems that the instrumentalist, too, owes us an explanation -– it can’t just be a \textit{miracle} that our theories are predictively successful.\cite{Putnam:1975NMA}  This "No-Miracles Argument" depends on the premise that the probability of a theory delivering predictive successes despite being false is very small (i.e. miraculous).

The rhetorical situation is interesting, because it seems that neither the instrumentalist nor the realist has an obvious answer to their respective challenges. And the status of the no-miracles argument itself remains controversial (see, for example, \cite{Howson:exhuming}) but the NMA highlights a plausible intuition: that we should take a theory more seriously when it makes accurate predictions that are difficult to get right.

What does this have to do with the explanatory power of neural network models?  Where the scientific realists were concerned about the truth of their theories, we are concerned that the components of a mechanistic neural network model can be mapped to entities involved in the target system or phenomenon, and that dependencies in the model correspond to real causal dependencies in the target.\footnote{See our companion paper on mechanisms \cite{cao2021explanatory} for the full discussion of this requirement.} So while modelers need not talk about their models being “true” or “accurate representations” (although they may choose to), they must deal with a structurally similar issue. 

In a defense of what he calls “constructive realism”, Giere describes a paradigm case of a difficult prediction problem –-- the one faced by Watson and Crick --– where getting the prediction right is excellent justification for supposing the truth of the hypothesis: 
\begin{quote}
“The realistic hypothesis, of course, is that the molecular structure of DNA is quite similar in the relevant respects to the proposed model. The evidence is the particular pattern of light and dark spots observed on X-ray photographs. 

Enough was known about the patterns produced by various molecular structures and about X-ray techniques to justify the conditional judgment that, if the hypothesis were correct, the specified pattern would very likely result. But the same body of background knowledge also justified the converse judgment. If the real structure were not very similar to the model, this particular pattern would be unlikely to result.”\footnote{Giere et al, p.192-4, Constructive Realism, from Science without Laws.~\cite{giere1999science}}
\end{quote}
 
We claim that we have good reason to suppose that predicting neural activity in V1, V4, and IT is an excellent example of a difficult prediction problem in Giere’s sense.  As with the X-ray diffraction patterns, the number of ways to get the predictions wrong dramatically outstrips --- by many orders of magnitude --- the numbers of ways to get the predictions right, within the error margins and to the pre-specified degree of approximation.

So we agree with the critic this far: in order for accurate prediction to provide independent evidence of mechanistic mappability, the prediction problem must be \textit{difficult}, as illustrated above.  But we disagree with the critic who thinks that in the case of ventral stream neural activity, the prediction problem is easy. We described some demonstrations of the difficulty in \S5 of the companion paper to this one, and see original references in the scientific literature~\cite{yamins:pnas2014,kriegeskorte:ploscb2014}. Additional non-trivial controls included attempts to predict neurons in each brain by: non-optimized neural networks (including shallower neural networks and all other previous models posited up to the time of that work), neural network layers corresponding to \emph{other} brain areas, neurons themselves from other brain areas, as well as ideal-observer models.  In each case, the mappings achieved significantly less predictive success that the correctly mapped layer of the task-optimized deep net.  If the prediction problem were easy, all these controls would not have come up short.

We've talked about how unexpected it would be that neural network models trained on nothing but labeled images manage to predict anything about the brain, if they did not \textit{also} share important functional features with the brain.  What makes it surprising is the exceedingly low probability that we might just \textit{chance} on a good prediction of the highly complex response patterns of monkey neurons in V1, V4 and IT to novel stimuli.  What makes successful prediction much more probable is that there is some genuinely deep similarity between the artificial NN models and the parts of the brain with which its capacities overlap, despite their very different origins.\footnote{In fact, the models that account for neural responses the best are those that were created by the most independent of sources: computer vision researchers who cared about winning image categorization contests, and not at all about neuroscience, and in fact hardly knew anything about it (even if they were using models inspired by people who were inspired decades earlier by neuroscience).}  The evolutionary perspective gives us a familiar framework in which to situate this outcome: that what we see is likely to be convergence to the best accessible solution, given the architectural and task constraints. 

\section{An Evolutionary Interpretation}

Thinking about evolution in the context of biological explanation helps us to make sense of particular organism features by understanding what they are \textit{for} --- what function they are supposed to perform, and how they contribute to fitness in a particular environment.  Similarly, if we understand the goal or function of neural activity as the production of the right behavior at the right time, we can make better sense of why particular neurobiological facts (functional, physiological, and anatomical) are as they are. And we've argued that when we focus on the process of quantitative optimization by which NN models are built, we can see \textit{why} they end up with the particular architectures and parameters that they do:\textit{because} of a sufficiently constraining goal that they were trained to optimize.

So we might add a supplementary observation to parallel Dobzhansky and Shepherd above: \textit{that nothing in computational neuroscience makes sense except in light of optimization}. What all three maxims have in common is that they point at something like a \textit{normative} function: a task to be performed, or a goal to be achieved.  In each case, the system --- the organism, the brain, or the neural network abstraction --- is \textit{teleonomic}; it was shaped --- by evolution, development, learning or training --- \textit{as if for a purpose}.  Whether that shaping is the result of natural selection, neural plasticity, or gradient descent, it is an optimization process resulting in a system that is effective at performing a task or function, as if it were \textit{designed} to be so. \footnote{This is not to claim that all (or even most) optimized systems are \textit{optimal} -- we can expect that artificial optimization or selection acting on a system will result in a reasonable level of performance on the goal or task, without presuming that optimal \textit{peak} performance is reached. So the brain may be optimized-for and therefore reasonably effective at many tasks, without necessarily being optimal at any of them.}

A core objective of neuroscience has been to uncover the underlying reasons why the structures of the brain are as they are, and why they respond as they do. Taking Dobzhansky's and Shepherd's dicta seriously requires us to identify the evolutionary and behavioral constraints that shape brain structure and function. That is, we want to know how function \textit{constrains} form --- and in so doing, see how function \textit{explains} form.\footnote{Note that if function did \textit{not} constrain form at all, then no such explanatory relation would obtain.}  It follows that any goal-optimized approach to neural network modeling has to take seriously the idea that organisms must actually produce useful behavior in the world --- that is, they must achieve task goals.

To the extent that that goal is \textit{shared} by NN models and the brain systems we are interested in explaining, elucidating the relationship between task success and mechanistic facts about the system successfully performing that task will help to explain both kinds of systems.  So we want to find tasks that are good proxies for evolutionary goals that brains were actually selected for achieving.

To make the parallel fully explicit, think of NN models as being built in the context of a tripartite optimization framework, consisting of an architecture, an objective function, and a learning rule. In the context of biological explanation, the objective function in an optimization framework may be replaced with some kind of fitness-related goal, the architecture with the physical anatomy of the organism, and the artificial learning rule with natural processes of evolution and ontogenetic learning.\footnote{While an oversimplification, the relationship between the optimization of discrete architecture parameters and synaptic strengths is somewhat analogous to that between developmental and evolutionary variation. The continuous changes to the synaptic strengths can be thought of as happening during development and learning and can occur without changing the overall system's anatomical structure.  Changes in the discrete parameters, by contrast, restructure the primitive operations available (e.g. in the cell-types present), the number of layers of processing (sensory areas) and the number of neurons in each area. Similarly, it may be conceptually useful to think about the outer loop optimization as occurring over evolutionary time, and the inner-loop optimization as occurring over life-time learning.}

We can think of an NN model as an individual under some kind of artificial selection. If this individual is to be a good model of a biological organism under natural selection, then its final output layers must be effective at performing the same high-level goal-oriented tasks that the organism had to succeed at, whatever the parameters the network ends up adopting. Structurally, the kinds of search through possibility space that the modeler undertakes are analogous to the kinds of search that result in competent adult brains shaped by evolutionary and learning and development processes.  In both, there are a large number of parameters, whose values are significantly shaped by non-convex optimization problems.  Since incrementally developing a brain with so many interesting and sophisticated capacities is itself a difficult task, subject to a multitude of constraints, perhaps we should not be so surprised that our two known solutions to one of those problems (that of visual object classification) were arrived at by structurally similar routes.

To the extent that evolution and neural network modeling are both optimization processes, optimization can tell us valuable things about evolved systems. The optimization approach can give us some evidence as to whether a task is constraining or not.  It can point us towards which features of the myriad physical facts about a system are functionally important for success at which tasks – potentially answering the question of why those features are present, and not others.  It may turn out that every physical detail is important for performance at some task, but we may not care about all tasks in all contexts, and optimization for particular tasks can tell us which things matter when.

The answers that optimization offers to why-questions will also have the same flavor as evolutionary explanations.  That is, they answer why-questions, what-for questions, or teleological questions, in much the same way as more familiar adaptationist explanations, such as efficient coding~\cite{chirimuuta2017explanation}.  So in fact, there is nothing \textit{new} in the mode of explanation that is offered by neural network models.  Biologists have \textit{always} appealed to facts about function: to take a salient example, claims about what different brain areas "represent" -- claims that are absolutely ubiquitous in neuroscience -- are in part claims about the function of the system. 

Now for some caveats. 

It is clear that many of the constraints faced by evolution are different from those faced by the neural network modeler.  Different materials mean different temporal constants and speed of processing.  Developmental constraints restrict possible architectures.  Metabolic constraints and other physiological constraints restrict energy use.  The repertoires of proteins, genes, and other biochemical parts restrict the sorts of structures that can be built.   And insofar as the evolutionary problem is different from the task posed by the modeler using their objective function, or the evolutionary solution comes from a completely different class of mechanisms (even when understood at an adequate level of abstraction) than that of the model architecture, the solution will also be different. 

\begin{figure}
\centering
\includegraphics [width=0.95\linewidth]{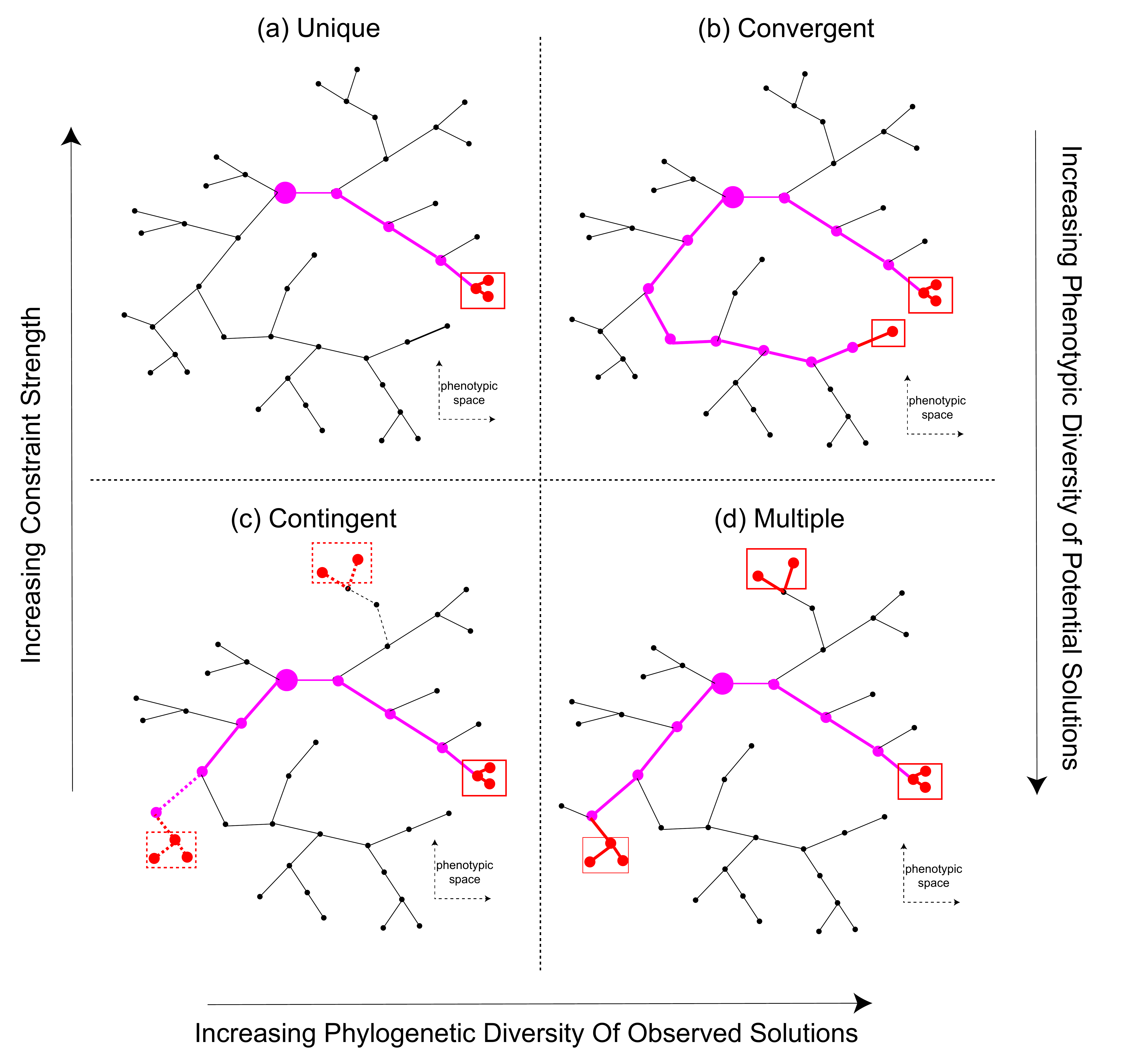}
\vspace{-3mm}
\caption{\textbf{Constraints on Evolutionary Paths.}  Each of the four phylogenetic trees represents a different possible evolutionary history, resulting from a different kind of interaction between the task constraint landscape and genetic history.  The red boxes represent clades in which the task constraint is met; the purple links represent the evolutionary path(s) arriving at such clades; and the large purple dot is the evolutionary common ancestor.  Panel \textbf{(a)} represents the situation in which only one possible phylogenetic path could ever have arrived at an organism with a phenotype that meets the task constraint.  Panel \textbf{(b)} represents the situation in which all solutions to the task must have similar phenotypes, arising in multiple quite phylogenetically distant organisms.  Panel \textbf{(c)} represents the situation in which there are multiple phenotypically-distinct solutions to the task that could have arisen genetically, but in which most never actually arose due to historical contingency, and only one of the possibilities is actually is observed. (The potential but not-actually-historically-observed possibilities are indicated by the dotted lines.)  In this situation, the genetic determinants of the observed solution might be conserved over long timespans, due not to the task constraint strength but rather path-dependency of evolutionary history.  Panel \textbf{(d)} shows the situation in which multiple phenotypically-distinct solutions actually arose.  We would expect models that solve the task to ``look like the brain'' in cases \textbf{(a)} and \textbf{(b)}, but not \textbf{(c)} and \textbf{(d)}.  
\label{fig:paths}}
\vspace{-3mm}
\end{figure}

We are \textit{not} claiming that evolution is always maximizing, or that natural selection explains everything in biology by itself. It may be that at some points in the phylogenetic tree of life, fitness goals were not very constraining, and so very different solutions were possible. It may be that much of evolutionary diversity and may be explained by neutral processes such as drift, rather than strong selective pressures. It may be that many features of the brain can be explained by developmental or path-dependent constraints on its form, rather than in-principle performance-based constraints of the kind that we are looking for good proxies for. (See Fig.\ref{fig:paths} for illustrations of these different scenarios). We agree that these issues are complex, and that there are many competing considerations (see for example \cite{okasha2018agents}, or \cite{Ohta16134}). Rather, we think that \textit{insofar} as some features of an interesting capacity exhibited in the biological world can be accounted for in adaptationist terms, an optimization-based model of that system will be helpful in explaining those very features -- and that we can use the optimization framework as a heuristic for hypothesis generation about likely features of naturally observed evolved systems.

We also know that biological systems produced by evolution and development are not guaranteed to be optimal for their evolutionary niche. But the same is true for the models produced by any practically implementable learning rule, being subject to the same problem that evolution/development faces: failures to achieve optimality due to incomplete optimization or capture by local maxima. The hope is that the failures of optimization for the real biological system are mirrored by those of the computational model -- and insofar as the model of the learning rule and distribution over initial conditions are themselves biologically accurate, the same patterns of performance failures should be observed in both the model and the real behavioral data~\cite{rajalingham2018large}.

As an approach to biological modeling, the optimization framework embodies a kind of methodological adaptationism.  It assumes that the biological facts to be explained are actually the result of strong functional selection for a stable evolutionary niche -- rather than a series of path-dependent random events or the result of changes during a time of extreme instability where no clear evolutionary niche obtained. This assumption is a pragmatic one: the alternative is that there is \emph{no} particular reason why a given structure is seen. In such a case, the optimization framework fails to give an explanation because in a certain sense, there \textit{is} no "why" -- the outcome was a mere accident of history.

One well-recognized way in which a system can be made intelligible is through the discovery of a concise mathematical description of its performance. We've now described an additional, distinctively biological way, in which a system can be made intelligible -- by showing how it results from the dual constraints of an ethologically plausible task and a biologically plausible mechanism. The more plausible the task and mechanism class, in turn, the more faithful the model is likely to be to the original target of explanation. 

As a matter of fact, we still observe finer-scale differences in performance between primates (with their evolutionary inheritance, trained in getting around in the world), and these most recent deep HCNN models (which we have designed, and then trained on databases we have constructed, to do somewhat artificial tasks that are similar but not \textit{identical} to evolutionarily specified goals)~\cite{rajalingham2018large}. But this is exactly what we should expect, if the heuristics delineated above are good ones. To the extent that we perform the same tasks, with the same resources, we should expect our solutions to be similar. To the extent that we are performing different tasks, with different constraints and resources, we should expect our solutions (and the underlying causal structure they represent) to be different. Again, the proof of the pudding is in the eating: the more accurate the predictions for a hard problem, the more likely it is that the tasks and constraints are actually similar.

\begin{figure}
\centering
\includegraphics [width=.95\linewidth]{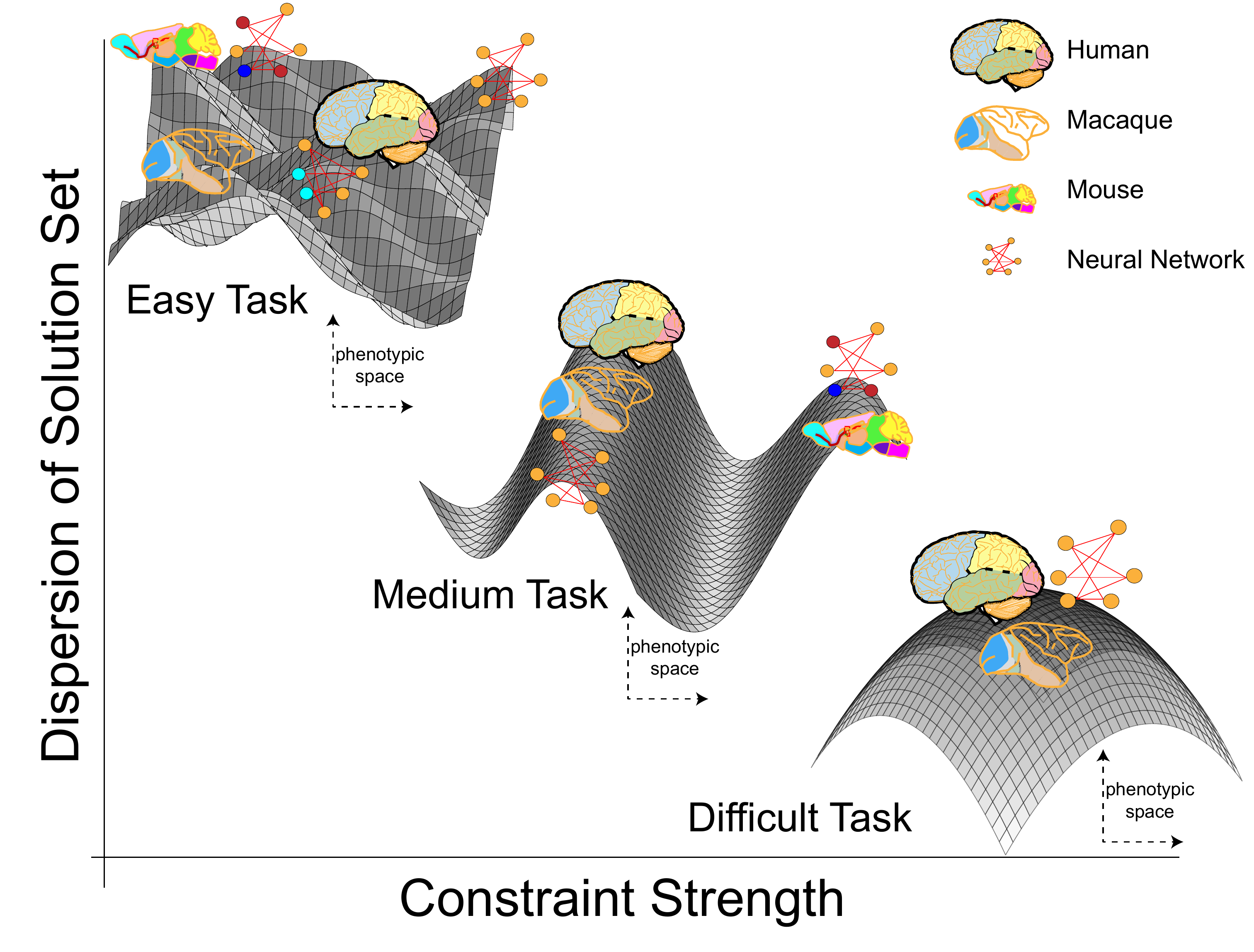}
\vspace{-3mm}
\caption{\textbf{The Contravariance Principle.}  Evolutionary constraints on the optimization target (the ``task''),  determine the shape of the landscape. Structural constraints on network architecture carve out a subset of the accessible landscape.
By ``dispersion'' of the solution set we mean: how large and variable the set of possible parameter settings corresponding to optimized solutions to the constraints can be.   
There is an inverse relationship between the strength of a constraint set and dispersion of the solution set, as mediated by the optimization landscape imposed by the constraints.  
The harder the task is, the stronger the constraint, and the more likely any given high-performing solution (e.g. a deep neural network) is similar to any other (e.g. an actual brain). 
The image depicts the situation as if there were a single evolutionary niche governing the evolution of all three species (human, macaque, and mouse), but this is unlikely to be true: instead, a more accurate (if unwieldy) picture would show an evolutionary tree of landscapes, with the actual niche pressures on each species defining its own landscape. Whether that species-relativized constraint landscape is actually sufficiently strong that it determines the architecture of solution fairly closely is an empirical-computational neuroscience question.  
\label{fig:contravariance}}
\vspace{-3mm}
\end{figure}

\section{The Contravariance Principle}
\label{subsec:contravariance}

The optimization approach can help us answer why-questions when there are strong dependencies between task and mechanism --- when function dictates mechanism.  But this won’t always be the case, so we want to specify the conditions under which we think the approach is best applied.

Though it may at first seem counterintuitive, the \emph{harder} the computational goal, the \emph{easier} the model-to-brain matching problem is likely to be.  This because the set of architectural solutions to an easy goal is large, while the set of solutions to a challenging goal is comparatively smaller.  In mathematical terms, the size of the set of optima or solutions is \emph{contravariant} in the difficulty of the optimization problem (see Fig. \ref{fig:contravariance}). 
 
A simple thought experiment makes this clear: imagine if, instead of trying to solve 1000-way object classification in the real-world ImageNet dataset, one simply asked a network to solve binary discrimination between two simple geometric shapes shown on uniform gray backgrounds.  The set of networks that can solve the latter task is much less narrowly constrained than that which solve the former.  And given that primates actually do exhibit robust object classification, the more strongly constrained networks that pass the same harder performance tests are more likely to be mappable (in the 3M++ sense) to the real primate visual system.  A detailed example of how optimizing a network to achieve high performance on a low-variation training set can lead to poor performance generalization and neurally inconsistent features is illustrated in \cite{hong2016explicit}. 

In general, let us define one task (call it $A$) as being harder than another task (call it $B$) if all solutions to $A$ end up also being able to solve $B$.  In this case, task $A$ places more or stronger constraints on the systems that are capable of successfully performing it than does $B$.  Tautologically, having more constraints (again, all else equal) means fewer systems that satisfy those constraints. That is, probability of success at the task is low, unless you have just the right set of features, arranged in just the right ways.  If it turns out that our object categorization task is difficult, then it is likely that any system that can successfully perform will have just the right set of functional subcomponents, arranged in just the right ways … and so any two systems that can do the task will be similar in functionally relevant ways (even if they may be dissimilar in functionally irrelevant ways).  We can see something similar in convergent evolution in nature --- the re-emergence of swimming fins in multiple lineages for example, which are physically similar in ways related to their function, even as they differ in their particular details and evolutionary histories.

Our fast, effortless, and accurate categorization of objects in natural scenes is plausibly a sufficiently difficult task to be highly constraining.  Our visual systems are at the same time highly selective (able to distinguish a huge variety of different objects), yet tolerant (to changes in retinal image generated by the same object due to differences in perspective, distance, and so on).  We think the problem is difficult, partly because we tried many things that failed before hitting on a set of solutions that performed relatively well.  (As Giere points out, scientists are rarely in the position of deciding between a multitude of equally good solutions --- rather, we are much more often in the position of struggling to come up with even one good model).  Moreover, independent paths --- one from evolution, the other from computer vision --- led to similar solutions.

Of course, there is no \textit{guarantee} that for a given neural system (or indeed any structure in biology) that it can be fully explained by strong task constraints, even when we only observe one (type of) solution in nature. As discussed earlier, there is always the possibility that even highly conserved structures (usually evidence for strong selection) were selected for primarily on the basis of path-dependent or accessibility constraints, rather than primarily task-related performance constraints.   That is, fitness might require only adequate performance at a not-very-constraining task, but limited genetic resources or rigid developmental constraints could still effectively reduce the space of solutions for that lineage.  Or perhaps only a single solution emerged in nature by mere historical contingency, even though other solutions were possible because the task was not difficult (see Fig. \ref{fig:paths} for illustrations of these different scenarios). In these cases, we should \textit{not} expect artificial models of the task that biological system performs to be strongly informative about the biological system.

This means that for modeling purposes, it is primarily the performance-based constraints that are exploitable to discover similarities between the operation of the biological system and our models, although they may be supplemented by developmental or other mechanistic constraints that can be built into the model (e.g. minimization of wiring length). So what makes a task a good proxy is not just that it is similar enough to the evolutionary goal that being good at one results in being good at the other, but also that that evolutionary goal itself was constraining enough -- over and above path-dependent constraints -- to force convergence to similar solutions in different systems.

The strategy we have described is a heuristic, one that does not work in all situations. We do not claim that there is a unique solution to any task.  But we do argue that given a challenging task, we should take seriously the possibility that two systems that solve it share deep explanatory similarities, even if those similarities are not immediately apparent. Just as difficult predictions justify realist interpretations, difficult tasks are more constraining tasks, and success at difficult tasks justifies mechanistic/causal interpretations of our successful models.\footnote{In words from the companion paper on mechanism, we should look for some PARA that describes both solutions.\cite{cao2021explanatory} Then, given that abstract description, we should look for 3M++-style mappings to real parts of the system, including (especially) those we did not build in ourselves. In this way, neural network models can potentially yield novel mechanistic hypotheses and insights.}

To put this explanatory approach into practice requires: postulating one or several goal behavior(s) as driving the evolution and/or development of a neural system of interest; finding architecturally plausible computational models that (attempt to) optimize for the behavior; and then quantitatively comparing the internal structures arrived at in the optimized models to measurements from large-scale neuroscience experiments. To the extent that there is a match between optimized models and the real data that is substantially better than that found for various controls (e.g. models designed by hand or optimized for other tasks), this is evidence that something important has been captured about the underlying constraints that shape the brain system under investigation. 

Though the contravariance principle is a tautology, it makes a strong and non-obvious prescription for using the optimization framework to design effective computationally-driven experiments.  Unlike typical practice in experimental neuroscience, but echoing recent theoretical discussions of task dimensionality~\cite{gao2017theory}, it does \emph{not} make sense from the optimization perspective to choose the most reduced version of a given task domain and then seek to thoroughly understand the mechanisms that solve the reduced task before attempting to address more realistic versions of the task.  In fact, this sort of highly reductive simplifying approach is likely to lead to confusing results, precisely because the reduced task may admit of many solutions that are not much like each other. "Fitting the data" would then be so easy that it would fail to distinguish different models that were all adequate to the task. 

Instead, it is more effective to impose the challenging real-world task from the beginning, both in designing training sets for optimizing the neural network models, and in designing experimental stimulus sets for making model-data comparisons.\footnote{This is not to say that experiments should not be as carefully controlled as possible, repeatable, etc. but rather that the task at the core of the experiment should be a difficult, ethologically plausible one. This might in turn make controls harder to implement, but that is a substantive trade-off worth weighing.}  Even if the absolute performance numbers of networks on the harder computational goal are lower, the resulting networks are likely to be better models of the real neural system.\footnote{Another way to constrain networks while still using a comparatively simple computational goal is to reduce the network size. Ultimately, there is a natural balance between network size and capacity.  In general, the optimization-based approach is likely to be most efficient when the network sizes are just large enough to solve the computational task. This idea is consistent with results from experiments measuring neural dynamics in the fruit fly, where a small but apparently near-optimal circuit has been shown to be responsible for the fly's simple but robust navigational control behaviors~\cite{turner2017angular}, although it remains unknown whether the specific architectural principles discovered in such simplified settings will prove useful for understanding the larger networks needed for achieving more sophisticated computational goals in higher organisms.}
    
Optimization gives teeth to evolutionary explanation, because unlike real evolutionary processes extending into the distant past to which we have sparse and indirect epistemic access, we have full access to the optimization process actually undergone by the models we build.  We can now test whether a particular constraint leads to a particular outcome, rather than merely speculating. By connecting function to form, and form to function, optimization-based models can help make sense of mechanism.

\newpage

\pagenumbering{gobble}
\footnotesize{
\bibliographystyle{naturemag}
\linespread{0.9}
\bibliography{refs}
}

\end{document}